\documentclass[review]{elsarticle}
\usepackage{hyperref}

\journal{Journal of \LaTeX\ Templates}

\biboptions{sort&compress}

\usepackage{graphicx}
\usepackage{float}
\usepackage{dcolumn}
\usepackage{bm}
\usepackage{xcolor}
\usepackage{amsmath}

\begin{document}

\begin{frontmatter}

\title{How the overlap of excluded volume determines the configurational energy landscape and ``thermodynamics" in the ``one to five hard disks in a box" system}

\author[mymainaddress,mymainaddress2]{Adri\'an Huerta\corref{mycorrespondingauthor}}
\ead{adhuerta@uv.mx}

\author[mysecondaryaddress]{J. Quetzalc\'oatl Toledo-Mar\'in}

\author[mysecondaryaddress]{Gerardo G. Naumis}
\cortext[mycorrespondingauthor]{Corresponding author}
\ead{naumis@fisica.unam.mx}

\address[mymainaddress]{Permanent address: Facultad de F\'isica, Universidad Veracruzana,
Circuito Gonz\'alo Aguirre Beltr\'an s/n Zona Universitaria,
Xalapa, Veracruz 91000, M\'exico}
\address[mymainaddress2]{Sabbatical leave: Departamento de F\'isica Cu\'antica y Fot\'onica, Instituto de
F\'{i}sica, Universidad Nacional Aut\'{o}noma de M\'{e}xico (UNAM),
Apartado Postal 20-364, 01000 M\'{e}xico, Distrito Federal,
M\'{e}xico.}
\address[mysecondaryaddress]{Departamento de Sistemas Complejos, Instituto de
F\'{i}sica, Universidad Nacional Aut\'{o}noma de M\'{e}xico (UNAM),
Apartado Postal 20-364, 01000 M\'{e}xico, Distrito Federal,
M\'{e}xico}

\begin{abstract}
In this work, the effects of excluded volume are studied in the ``one to five hard disks in a box" system. For one and two disks 
in different types of cages, the attractive and repulsive forces are calculated analytically. Attractive forces are due to excluded 
volume overlap, and thus allows to understand that in hard-core systems, second neighbors have an effective interaction, a fact that
is usually neglected when considering only collisions with first-neighbors. The same effects are observed for numerical
computations with five-disks in a box, as the distributions
of distances between disks and disks to walls suggest. The results indicate that first and second-neighbor excluded volume interactions 
are essential to determine the energy landscape topology. 
This work supports previous observations that suggests that second-neighbor
excluded volume interactions in hard-core systems are related with phase transitions.
\end{abstract}


\end{frontmatter}


\section{Introduction}

The study of monodisperse hard disks, hard spheres as well as hard-spheres with angular interaction and polydisperse systems has a long history, yet still being the focus in many research arenas \cite{alder1962, wang1965significant, hoover1972exact,hoover1979exact, kawamura1979simple, kawamura1980simple,speedy1999glass, santen2001liquid, kern2003fluid,ashwin2009complete, wang2012colloids, smallenburg2013liquids,russo2017disappearance}. The state-of-the-art of such study has mainly been driven by open problems in statistical mechanics and soft matter such as the liquid-crystal transition \cite{kawamura1979simple}, the glass transition \cite{dyre2006col}, the liquid-liquid transition \cite{palmer2018advances} and supercritical liquid \cite{brazhkin2013liquid}, to name a few. Rigidity and relaxation \cite{phillips1979topology, thorpe1983continuous} are among the main features which still are not entirely understood. For this reason and in spite of the great evolution in this field, monodisperse hard-disks still provide the necessary ingredients to understand the freezing-phase-transition without having to deal with more than one parameter.

 In this sense, it is important to stress contributions ranging from theoretical approaches and computer simulations done by Alder and Wainright \cite{alder1962}, Eyring {\it et. al.} \cite{wang1965significant} as well as Hoover {\it et. al.}  \cite{hoover1972exact,hoover1979exact} and Kamamura \cite{kawamura1979simple, kawamura1980simple}.
Many years ago, the free volume concept was introduced as the available space that a particle has  with its neighbors held fixed. Such concept  plays an important role in the understanding the fluid motion and its structure, \cite{hoover1972exact,hoover1979exact}. Hoover  {\it et. al.}  pointed out that this structure has been well characterized by molecular dynamics and Monte Carlo simulations. On the one hand, Monte Carlo simulations have shown that from the time-averaged point of view, the free volume fluid structure is basically correct since the method chooses particles at random, meanwhile the remaining particles are held fixed, sampling the free volume and producing single-particle ``cells" due to the excluded volume of the fixed particles.

Hoover, et al., \cite{hoover1979exact, hoover1972exact} pointed out that as the density of the system increases in the fluid state, cages are created due to the overlap of excluded volume, decreasing in this way the available free volume of each particle to move. They have shown that a percolation transition, where the free volume changes from extensive to intensive, ocurrs at quite low densities about one-third of the freezing density, \cite{hoover1979exact, hoover1972exact}. As the density increases, the free volume decreases and the overlap of excluded volume decreases the mobility of the particles, giving place to the collective motion of the particles to sample the remaining available free volume.
  

More recently Huerta, et al, \cite{huerta2006freezing}, have shown a connection  with the second oscillation splitting formation of the radial distribution function, which is equivalent to the overlap of excluded volume with the second nearest neighbors, representing a more strict caging mechanism that produce the freezing transition in a hard disk system. Using that idea and a cell theory, a Van der Waals-like loop in the equation of state appears, which qualitatively agree with simulation results of the pure hard-disk systems. More interesting is the appearance of collective transverse modes near the split of the second oscillation of the radial distribution function, \cite{huerta2015collective}.

Thus, the previous results in fact suggest that the excluded volume overlap has a profound role in the dynamics and thermodynamics of the system, specially for driving phase transitions. Here, in order to understand the role of the overlap of excluded volume, we have studied the forces, distribution of distances between particles and particles with the walls of the one to five hard disk models in a box model. This allows to quantify the restrictions on the motion of the particles due to such overlap and the consequent formation of the configurational energy landscape.  Our results complements the early ideas of Hoover {\it et al.} \cite{hoover1972exact} who showed that for a hard-disk gas in the diluted limit, the overlap with the first-neighbor does not occur.

In fact, years ago Bowles and Speedy \cite{bowles1999five} showed that a system composed by only five hard disks confined inside a 2D hard box present states that are analogues to that of a fluid, crystal, supercooled and glass phases of a much larger system. They have described the behavior of the model in terms of the configurational landscape, this having five basins, namely one which represents crystalline state and four which represent the amorphous-like configurations \cite{bowles1999five}. When a hard disk configuration, either crystalline or amorphous, is highly confined, the particles can not dynamically change this configuration, the restrictions only allow vibrations of the particles inside their own cage produced by the neighboring particles and the hard walls of the container box. Meanwhile when the size of the box allows dynamical rearrangement of the particles, some of them can leave their cages and the system can explore different basins. Other finite size system have been studied in order to understand the mechanisms that drive them to the solid state. For example, Awazu as well as Speedy have studied systems composed by only two particles, \cite{speedy1994two,awazu2001liquid}, surprisingly these systems also present some features that gives some light in understanding how the caging and uncaging mechanism works on the relaxation process and thermodynamic properties of larger systems. A similar caging and uncaging mechanism has been analyzed more recently by Sirono \cite{Sirono}, in the context of glass formers with two step relaxation of a binary mixture of hard disks. Also, this two-disks system has been used to test hopping rate-theory between basins in the energy landscape \cite{bowles2004calculating}

Motivated by these ideas, here we investigate how such second-neighbor excluded volume arises in few disks systems, as the landscape is simple to understand\cite{bowles1999five}. Thus we start with a general framework on how overlap excluded volume can be treated for small systems in order to obtain effective forces. Then we apply such ideas to one and two disks in a cage. Then we study the case of five disks in a box, and finally, the conclusions are given. 


\section{Excluded area overlap and thermodynamics}

To understand the role of excluded volume overlap on the thermodynamics of the system, we can proceed as follows. First we consider general ideas of the relationship between excluded area overlap. Then we apply these ideas to small systems, i.e., a caged particle in a polygonal box and two particles in a box. The former case is important as the solid is well approximated by this.

\subsection{Excluded area overlap, equations of state and forces}

\begin{figure*}[htbp] 
	\centering
	a)\includegraphics[scale=0.20]{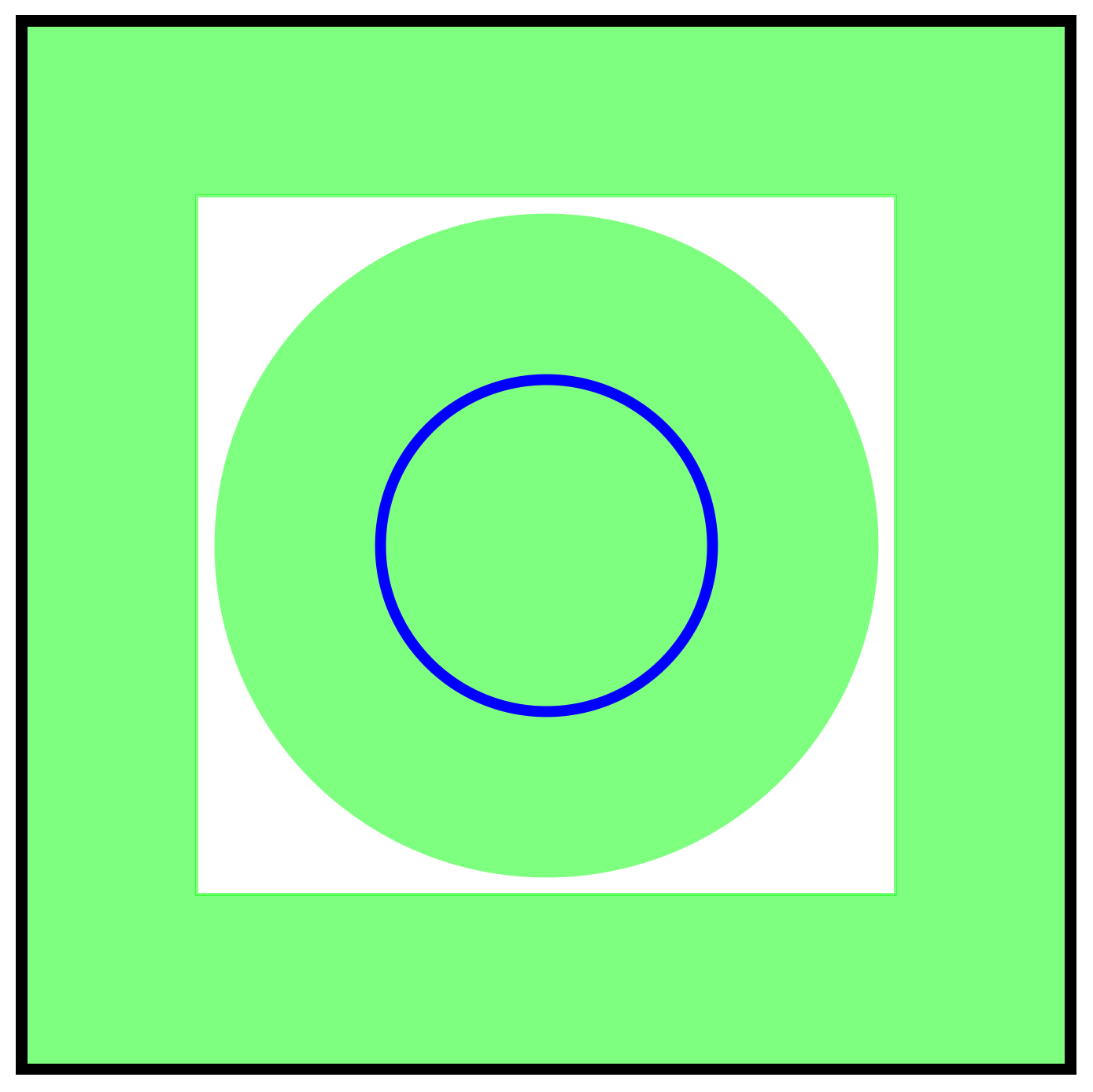}	  
	b)\includegraphics[scale=0.20]{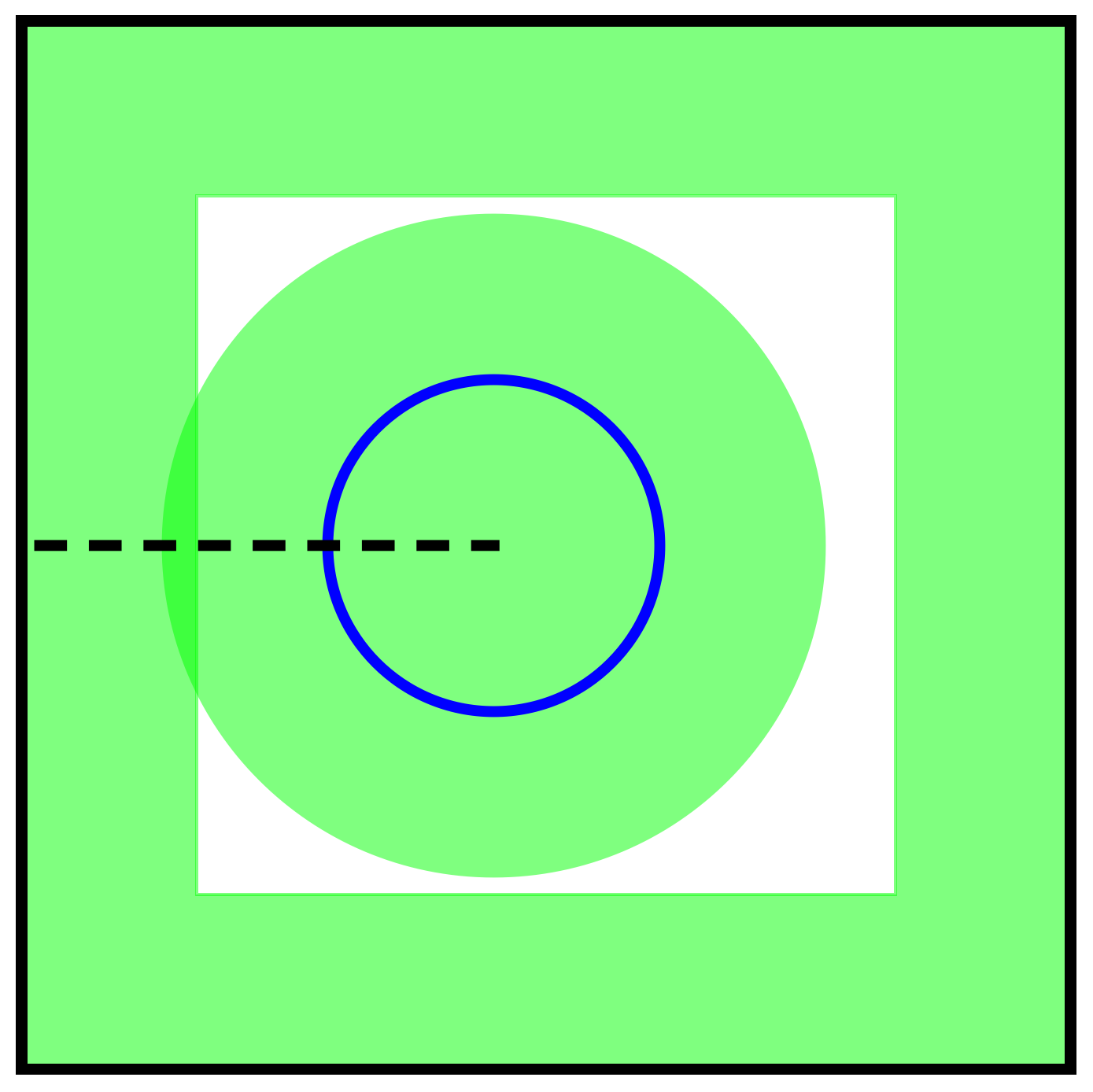}	  
	c)\includegraphics[scale=0.20]{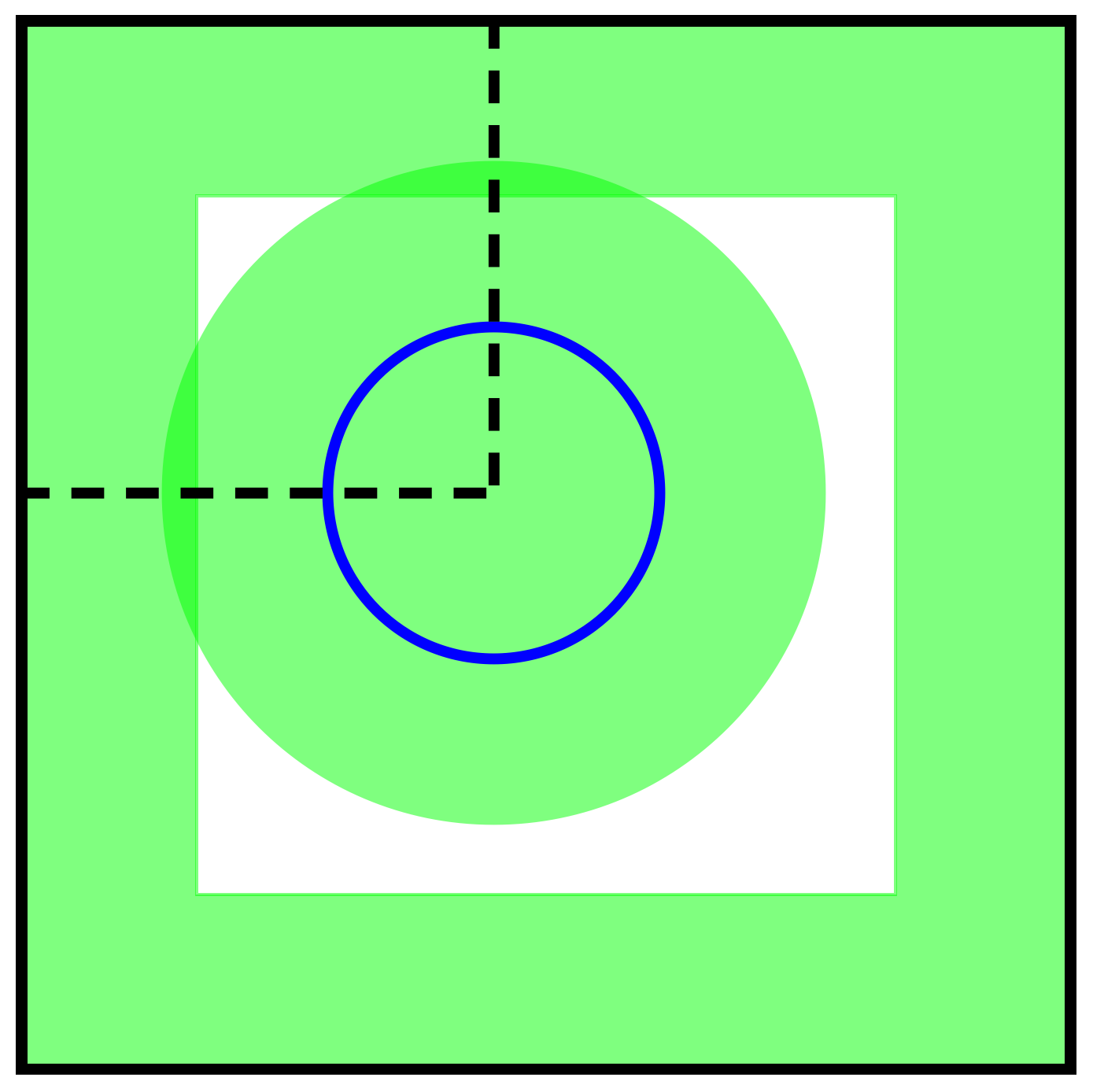}	   
	d)\includegraphics[scale=0.20]{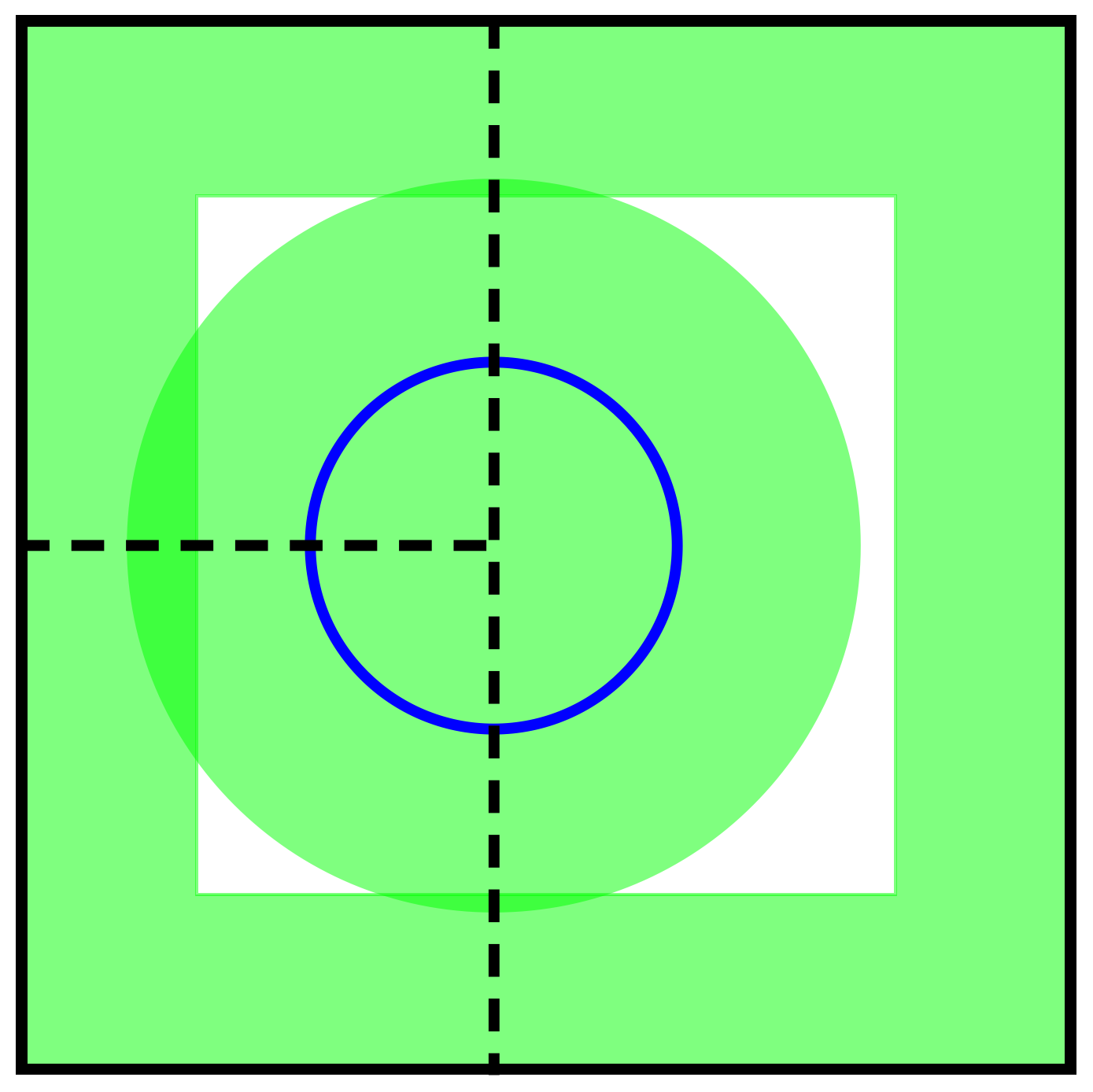}	
	e)\includegraphics[scale=0.20]{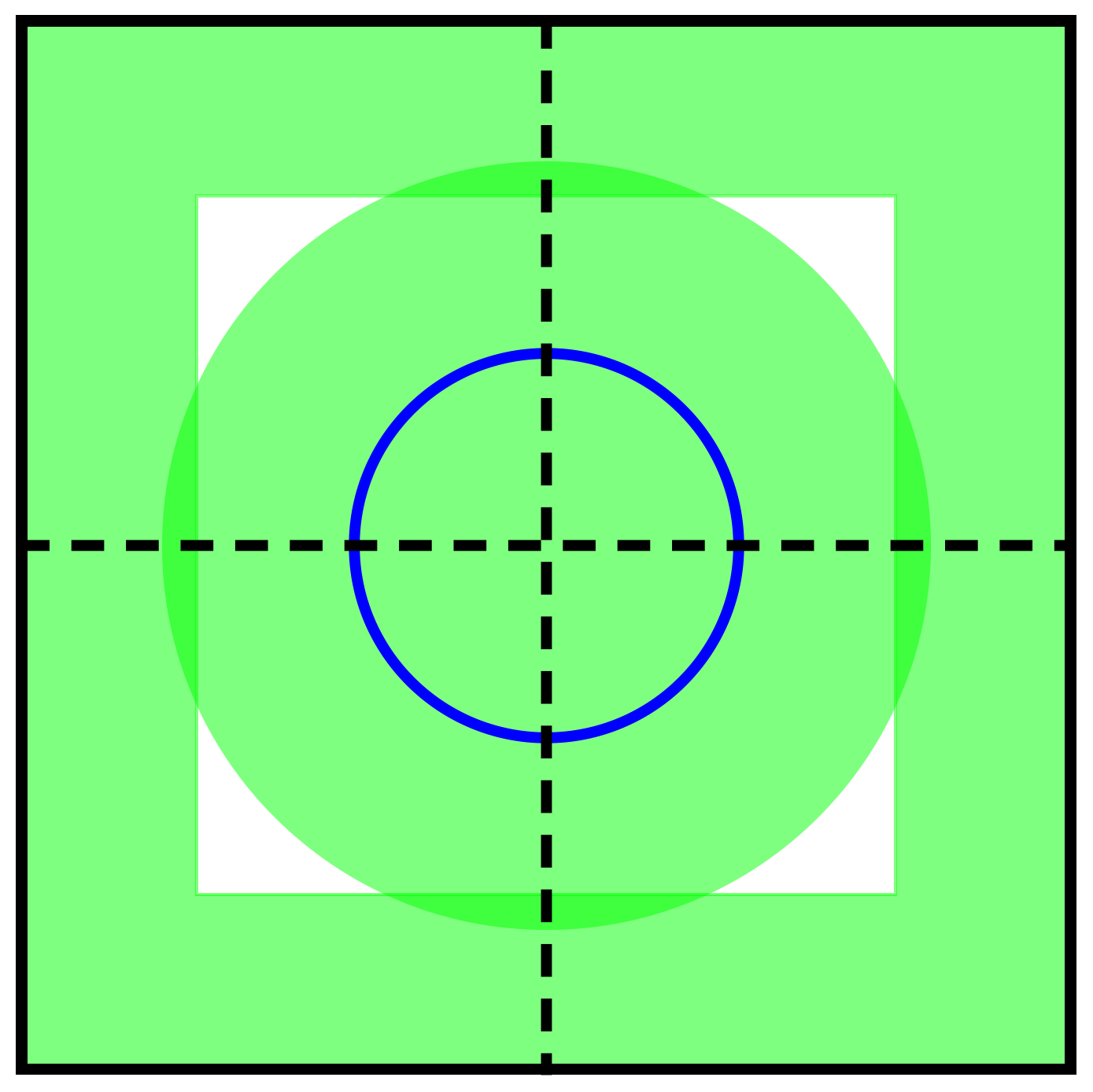}
	
	\caption{Snapshots of one hard disk in a box. In blue we show the disk of diameter $\sigma$ while in light green we show the excluded volume. We also indicate in green the excluded volume due to the hard walls of the box. The dashed lines in black stand for the overlap of excluded volumes between a wall and the particle, the dark green shows this overlap.}\label{Fig:snapshot1box} 
\end{figure*}

Let us start by considering a microcanonical ensemble for one hard disk ($N=1$) with diameter $\sigma$ in a box of area $A$. The box has side length $L$. As depicted in Fig. \ref{Fig:snapshot1box}, for the disk indicated in blue the number of accessible states is given by $\Gamma(1,A)=A_c$, where $A_c=(L-\sigma)^{2}=A+\sigma^{2}-2\sigma A^{1/2}$. This is the accessible space for one disk due to the wall constraints. These constraints define the green area next to the box walls, as seen in Fig. \ref{Fig:snapshot1box}. Observe that $A_c$ contains an overlap of excluded area (which in what follows we will refer to as the excluded volume, following the  literature practice)  of the disk and box walls. Each overlap is indicated by darker green and further emphasized by dashed black lines in Fig. \ref{Fig:snapshot1box}.

When a second disk arrives to the box, the accessible area is no longer $A_c$, as we need to exclude the area taken by the first disk, which is delimited in Fig. \ref{Fig:snapshot1box} by the green circle of the particle as well as the green bands of the box. Usually, in a very dilute system this area is approximated by  $\Gamma(1,A)\approx A(A-\pi \sigma^{2})$. A better approximation is obtained by observing that the first disk can be put in the area $A_c$, and then  $\Gamma(2,A)\approx A_{c}(A_{c}-\pi \sigma^{2})$. Even at this simple level, we can understand that the absence of excluded volume of the disk and walls overlap is the missing ingredient here. Thus, we need to correct this over counting. In fact, all incorrect counts correspond to the areas indicated by dashed lines in  Fig. \ref{Fig:snapshot1box}. It is essential to observe that even in Fig. \ref{Fig:snapshot1box},  the percolation of excluded volume overlap indicates a situation in which the available volume is separated in basins. 

\begin{figure*}[hbtp] 
	\centering
	a)\includegraphics[scale=0.20]{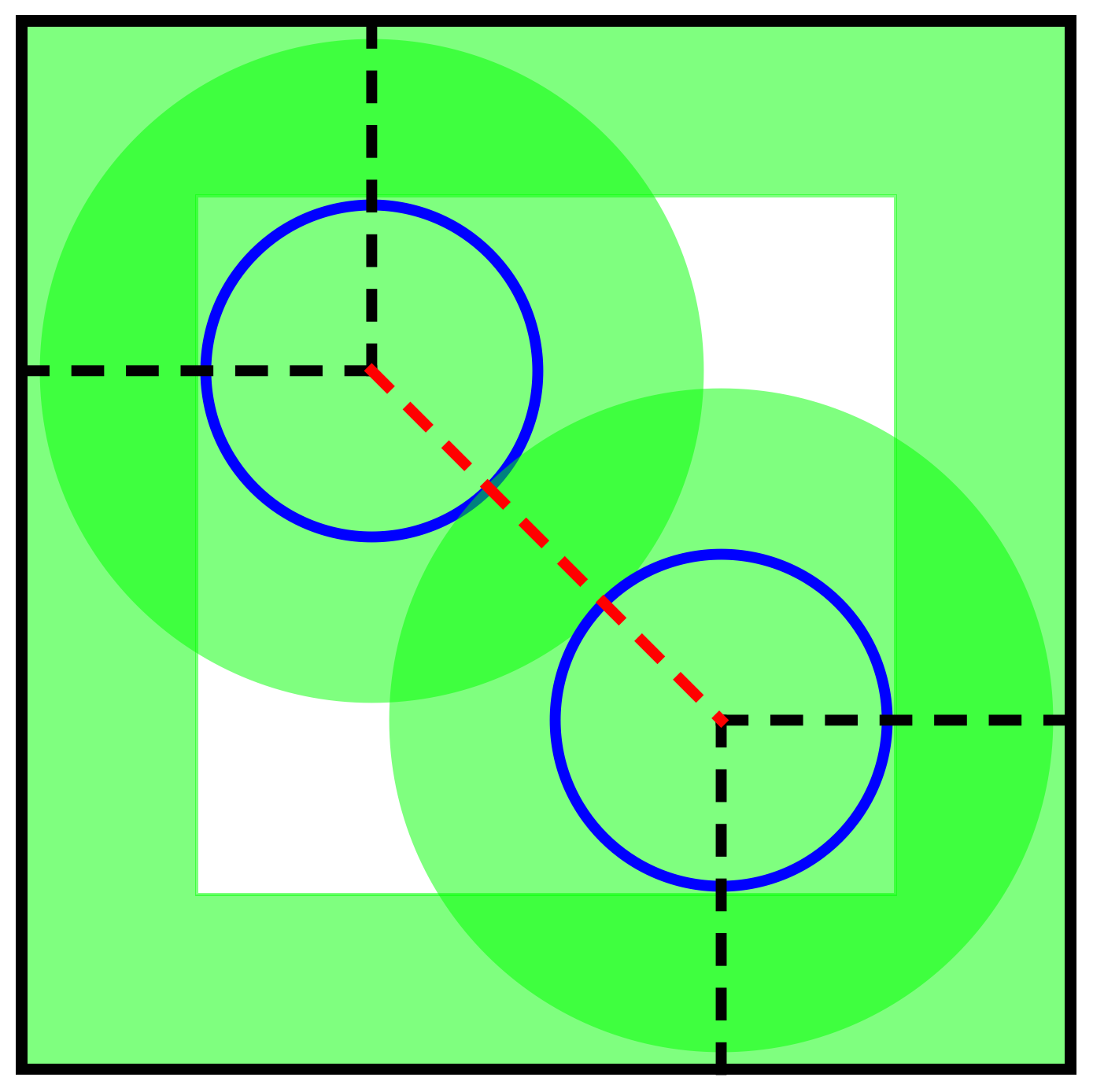}	  
	b)\includegraphics[scale=0.25]{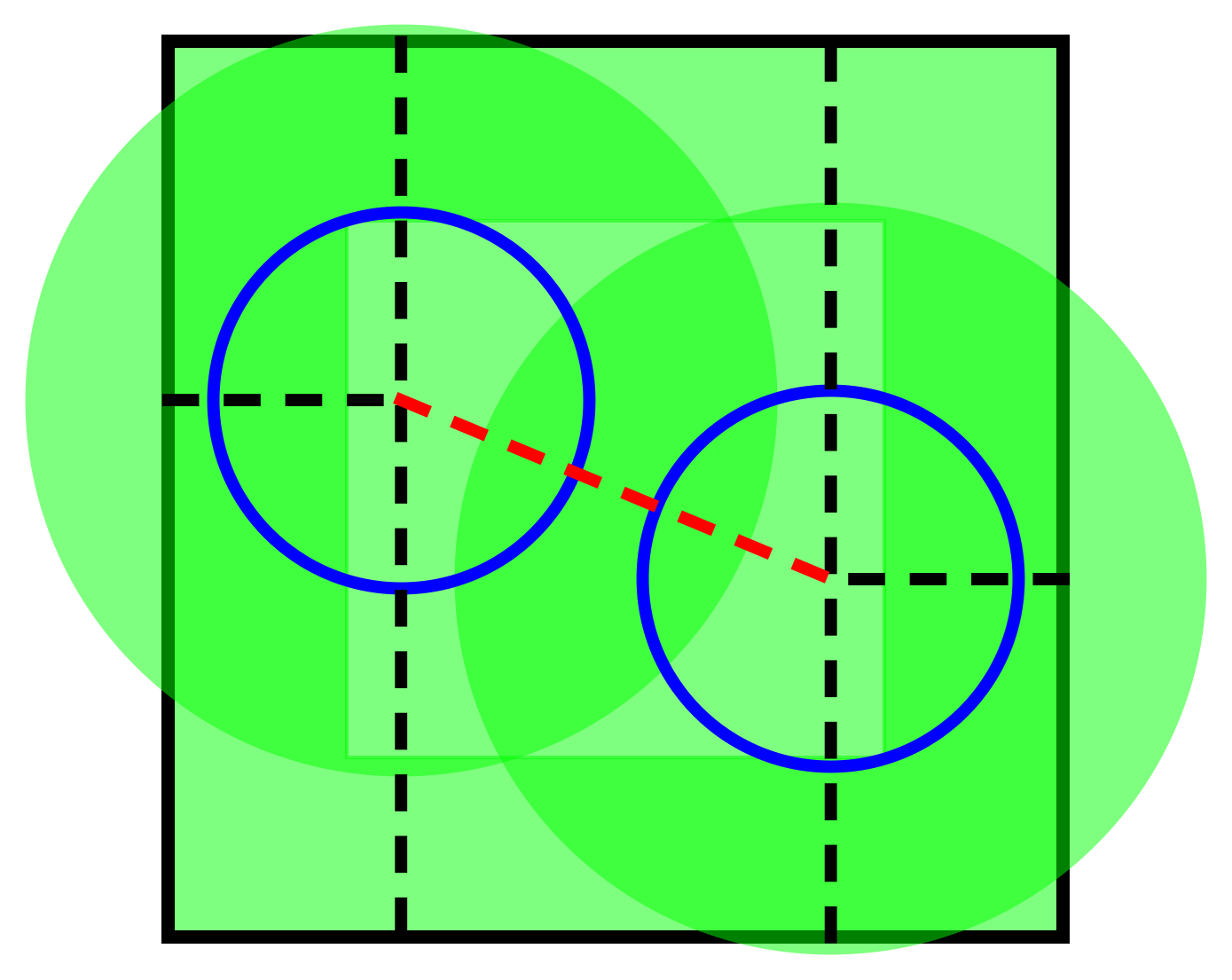}	  
	
	\caption{Snapshot of two hard disks in a box. In blue we show the disks of diameter $\sigma$, in green we show the excluded volume of each hard disk, also in green the excluded volume with the hard walls of the box. As in the previous figure, the dashed lines in black stand for the overlap of excluded volumes between a wall and the particles, whereas the red lines stand for the overlap of excluded volumes between particles, which are shown in darker green.}\label{Fig:snapshot2box} 
\end{figure*}

When a third particle is added, we need to look at the available free volume left by the other two particles. In Fig. \ref{Fig:snapshot2box} we show several snapshots of the situation.   The count   $\Gamma(3,A)\approx A_c(A_c-\pi \sigma^{2})(A_c-2\pi \sigma^{2})$ is incorrect as the excluded volume of the first-disk and the walls are counted twice when they intersect.

Thus, in principle we can make an exact calculation of $\Gamma(N,A)$ by including at each step the fraction $f_j$ of effective overlap between excluded volume,
\begin{equation}\label{Eq:GammaExact}
\Omega(N,A)= A_c^{N}\Pi_{j=1}^{N-1}\left(1-\frac{j \pi \sigma^{2}}{A_c}(1-f_j)\right).
\end{equation} 
where $f_j$ is yet an unknown function of the area $f_j=f_j(A)$ which contains the required overlap overcounting  at each step. For $N=1$, we can define an $f_0$ in such a way that, $A_c=A(1-f_0)$ its value is obtained from the wall and particle excluded area (see next subsection). Now, Eq. (\ref{Eq:GammaExact}) and $f_0$ allows us to get the equation of state from the entropy $S=k\ln(\Omega(N,A)/N!)$,
\begin{equation}\label{Eq:PExact}
\frac{P}{\rho kT}=1+\frac{\rho}{(1-f_0)}\frac{df_0}{dA}-\frac{4\eta j\frac{d}{dA}\left[\frac{1}{A}(\frac{1-f_{j}}{1-f_0})\right]}{\sum_{j=1}^{N-1} \ln \left[1-\frac{j \pi\sigma^{2}}{A}(\frac{1-f_{j}}{1-f_0})\right]}
\end{equation}
The second term in the previous Eq. contains the box contribution. For periodic boundary conditions $f_0=0$. The third term in the previous Eq. contains the excluded area and excluded overlap area contributions.

Finally, for a square box, we can obtain the normal force, $F$, which for here on we refer to as force,  using the fact that $dA=2LdL$ and that the force equates with the change rate of the configurational entropy by
\begin{equation}\label{Eq:force}
F=T\left(\frac{\partial S}{\partial L}\right)_N=2TL\left(\frac{\partial S}{\partial A}\right)_N
\end{equation}

Now let us ask about what are the advantages of writing Eq. (\ref{Eq:GammaExact}) and Eq. (\ref{Eq:force}) by using the overlap of excluded volume. There are several listed below:
\begin{itemize}
 \item It shows that second neighbors can interact with a disk, i.e., any disk in the range $\sigma<r< 2 \sigma$ has an effective interaction. Although effective interactions have been considered previously for hard-core systems, usually only first-neighbors are considered.
 \item Eq. (\ref{Eq:GammaExact}) and Eq. (\ref{Eq:force}) predict  attractive and repulsive forces.
 \item Repulsive forces are always bigger as thermodynamical stability requires a positive isothermal compressibilty.
\end{itemize}

\subsection{One disc in a polygonal box}

In order to apply the previous ideas, in this section we consider a particle in a polygonal box with $n$ sides. This serves as an approximation for a solid and as the simple test of the ideas developed in the previous section. 

We start by observing that the area of the $n$-polygon is $A=nD^{2}\tan(\pi/n)$, where $D$ is the distance between the center and the middle point of one of the any sides, i.e., the polygon apothem. Not all the area is accessible to the disk due to the excluded volume, resulting in the effective area $A_c=n(D-\sigma/2)^{2}\tan(\pi/n)$. Using that $D=\sqrt{A/n\tan(\pi/n)}$ we obtain,
\begin{equation}\label{Eq:One}
A_c=A-a_1\sigma A^{1/2}+a_0\sigma^{2}
\end{equation}  
with,
\begin{equation}
\begin{cases}
a_0= n \tan(\pi/n)/4 \; , \\
a_1=\sqrt{n \tan(\pi/n)} \; .
\end{cases}
\end{equation}
By setting $A_{c}=A(1-(1-f_1)\pi \sigma^{2}/A )$, we obtain the fraction of overestimated area due to the box,
\begin{equation}
f_0=-a_1\frac{\sigma}{A^{1/2}}+a_0\frac{\sigma^{2}}{A}
\end{equation}
.

Now we proceed to find the equation of state, as $\Gamma(1,A)=k_{B}\ln A_c$, from where it follows,

\begin{equation}\label{Eq:StateOne}
\frac{P}{\rho kT}=\frac{1-\left(\frac{\eta}{\eta_n}\right)^{1/2}}
{1-2\left(\frac{\eta}{\eta_n}\right)^{1/2}+\left(\frac{\eta}{\eta_n}\right)}
\end{equation}
where, 
\begin{equation}
\eta_n=\frac{\pi}{n \tan(\pi/n)}
\end{equation}

\begin{table}
\centering
	\begin{tabular}{lllllll}
			\hline
		$n$ & $3$ & $4$ & $5$ & $6$ & $7$ & $\infty$  \\ \hline
	$\eta_{n}$ & $0.6045$ & $0.7854$  & $0.8648$  & $0.9069$  &  $0.9319$ & $1$  \\ \hline
	\end{tabular}
 \caption{Maximal packing fractions of an $n-$sided polygonal box. $n=\infty$ denotes a circle. \label{tab:etas}}
\end{table}

Using Eq. (\ref{Eq:force}), we can obtain the force $F$ considering the length of the cage $l=2D$ as twice the polygon apothem. Notice that for even-sided polygons, $l$ coincides with the distance between parallel sides. Following Speedy \cite{speedy1994two}, we use the dimensionless variable,
\begin{equation}
 z= \frac{2\sigma}{l} \; .
\end{equation}

The resulting force can be written as, 
\begin{equation}\label{force2}
F=F_A+ F_R \; ,
\end{equation}
which has a repulsive, $F_R$, and an attractive, $F_A$, component, namely,
\begin{eqnarray}
F_R&=& \sigma k_B T \frac{z}{\left( 2-z\right)^{2}} \; ,  \label{eq:1diskForceA} \\
F_A&=& - \frac{z}{2} F_R\; . \label{eq:1diskForceB}
\end{eqnarray}

\begin{figure}[hbtp]
\centering
\includegraphics[width=3.3in]{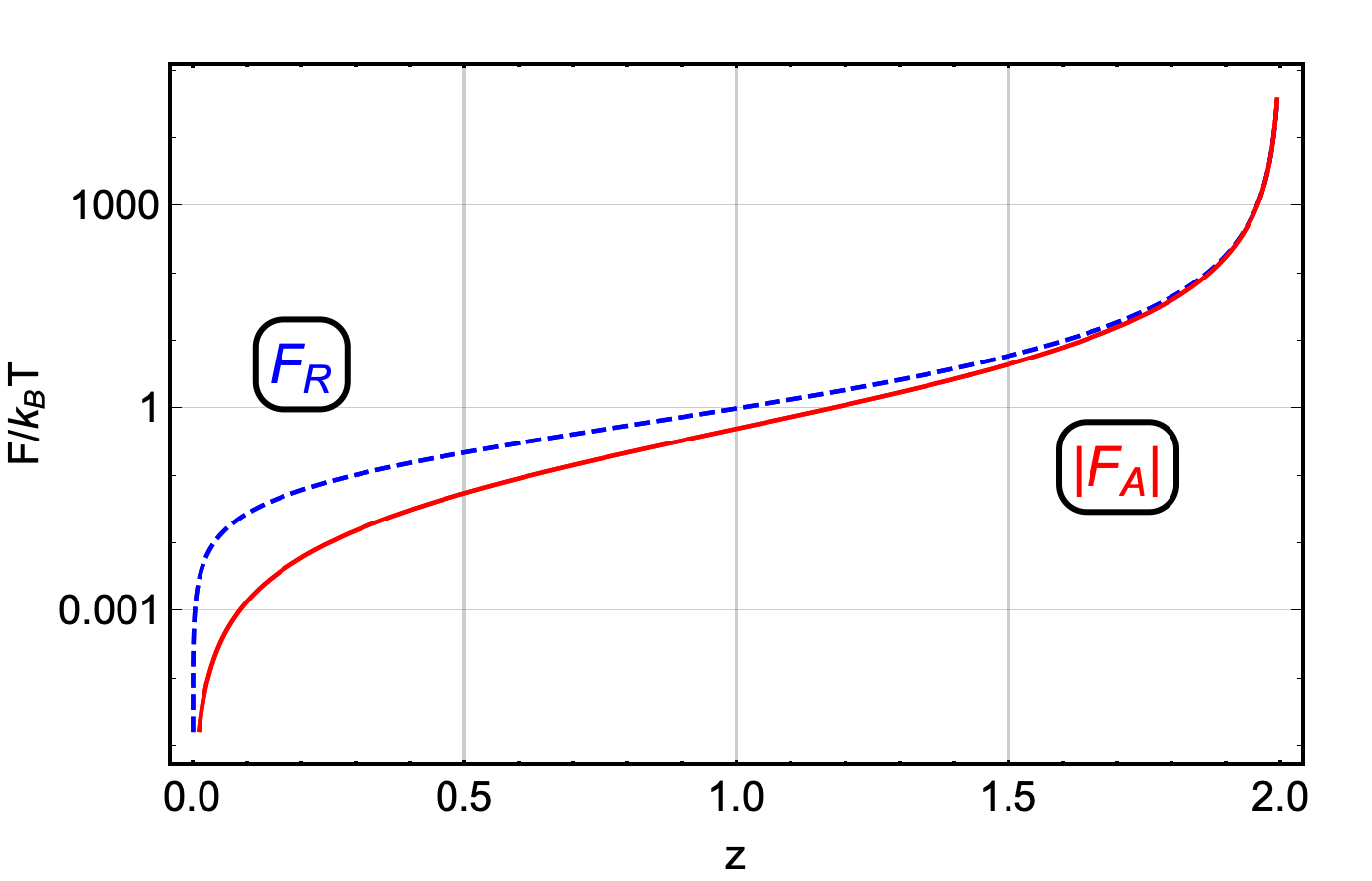}
\caption{Magnitude of the repulsive,  $F_R$, and attractive,  $F_A$, forces in the case of a disk in a box, given by Eqs. (\ref{eq:1diskForceA}) and  (\ref{eq:1diskForceB}), respectively. The curves are for a disk in a polygonal box. Notice that $\vert F_R \vert > \vert F_A \vert$ for all values of $z$. For $z \rightarrow 2$, $\vert F_R \vert \approx \vert F_A \vert$ as the excluded volume overlap is nearly equal to the excluded volume.} \label{fig:forces1Disk}
\end{figure}

In Fig. \ref{fig:forces1Disk} we have plotted the forces in Eqs. (\ref{eq:1diskForceA}) and (\ref{eq:1diskForceB}). Notice that  $\vert F_R \vert > \vert F_A \vert$.  As $z \rightarrow 2$, $\vert F_R \vert \approx \vert F_A \vert$ since the excluded volume overlap is nearly equal to the excluded volume. In general  $\vert F_R \vert > \vert F_A \vert$ due to the requirement of thermodynamical stability. When  $\vert F_R \vert \rightarrow \vert F_A \vert$ the system is close to a phase transition.

Several remarks arise from this example. The first one is that Eq. (\ref{Eq:One}) contains three contributions which are the area $A$, the interaction corresponding to the excluded volume overlap between the box and disk, going as $\sigma A^{1/2}$  and the excluded volume of the disk itself, of order $\sigma^{2}$. This is a general feature that will also appear for $N>1$. As $A_c \rightarrow 0$, these overlap terms begin to count over $A$, leading to a change in the scaling behavior of the system. In fact, notice that in Eq. \ref{Eq:StateOne}, $P$ diverges as $\eta \rightarrow \eta_{n}$, the divergence goes as $\sim A^{1/2}$.

We expect Eq. (\ref{Eq:StateOne}) to serve as a coarse-grained approximation to a solid or a dense fluid as particles are caged.  For $n=6$, a disk is caged in a effective closed hexagonal box. However, in a solid the cage sides are not straight lines but rather arcs of circumferences, as they are formed by surrounding disks. In spite of all approximations, from Table \ref{tab:etas} one would expect a phase transition between $\eta_3=0.6045$ and $\eta_4=0.7854$ since $4$ contacts on average are required  to have a rigid system and $n=3$ is the smallest sided polygon. Freezing for hard-disks occurs at $\eta\approx 0.69$ and melting at $\eta \approx 0.73$. In fact, a theory based on caging due to Wang, Ree, Ree and Eyring leads to a good equation of state for the solid \cite{wang1965significant}, while a triangular-box theory using next-nearest  neighbors gives an excellent estimate of $\eta=0.690$ for freezing \cite{huerta2015collective}. Finally, it is worthwhile to observe why the virial theorem is not able to reproduce a dense fluid or a solid, as the equation of state contains fractional powers of $\eta/\eta_n$, not obtained from an expansion with only integer coefficients.

\subsection{Two discs in a box}

Let us now consider the case of  two disks inside a square box of length $L$ with  {\it periodic boundary conditions}, i.e., $f_0=0$. Here there are two cases. If the length of the box $L$ is such that $L>2\sigma$, the 
first  disk can move  over  the whole area $A=L^{2}$ and, irrespective  of  its position, the second  disc has an available space $A-\pi\sigma^{2}$. Whenever  $L<2\sigma$, there is an excluded volume overlap $f_1$, as the first disc images on neighboring cells overlaps with each other. Thus, the first disc has an accesible area $A$, while the second has $A-\pi\sigma^{2}+A_o$, where $A_o=f_1\pi\sigma^{2}$ is the total excluded area overlap between images of the first disk. Using the results by Speedy \cite{speedy1994two}, we finally find that,
\begin{equation}
\Omega(2,A)=A^{2}\left[1-\frac{\pi \sigma^{2}}{A}(1-f_1) \right] \; ,
\end{equation} 
where the fraction of excluded overlap is,
\begin{equation}
\begin{cases}
f_1=0,  \; A>4\sigma^{2}\; . \\    
f_1=\frac{1}{\pi}\left[ 4\sec^{-1}\left( \frac{2 \sigma}{A^{1/2}} \right)+\frac{A}{\sigma^{2}}\left( \frac{4 \sigma^{2}}{A}-1 \right)^{1/2} \right] \; , \; A<4\sigma^2 .
\end{cases}
\end{equation}
Thus, the force will also have a repulsive and an attractive components, where
\begin{equation}
F_R=
\begin{cases}
\frac{2k_B T}{\sigma} \frac{z}{1-\frac{\pi z^2}{4}} \; ,z<1\; . \\
\frac{2k_B T}{\Omega(2,z)} \left(\frac{16  \sigma^3}{z^3} \left(1+\sqrt{z^2-1} \right) + \frac{6 \sigma^3}{z} \sec^{-1} \left(z \right) \right) \; , z>1\; .
\end{cases} \label{eq:2diskForceB}
\end{equation}
and 
\begin{equation}
F_A=
\begin{cases}
-\frac{2k_B T}{8 \sigma} \frac{z^3 \pi }{1-\frac{\pi z^2}{4}} \; , \text{for} \; z<1\; . \\
-\frac{2k_B T}{\Omega(2,z)} \left(\frac{2\pi \sigma^3}{z} + \frac{6 \sigma^3}{z \sqrt{z^2-1}} \right) \; , \text{for} \; z>1\; .
\end{cases} \label{eq:2diskForceA}
\end{equation}

\begin{figure}[hbtp]
\centering
\includegraphics[width=3.3in]{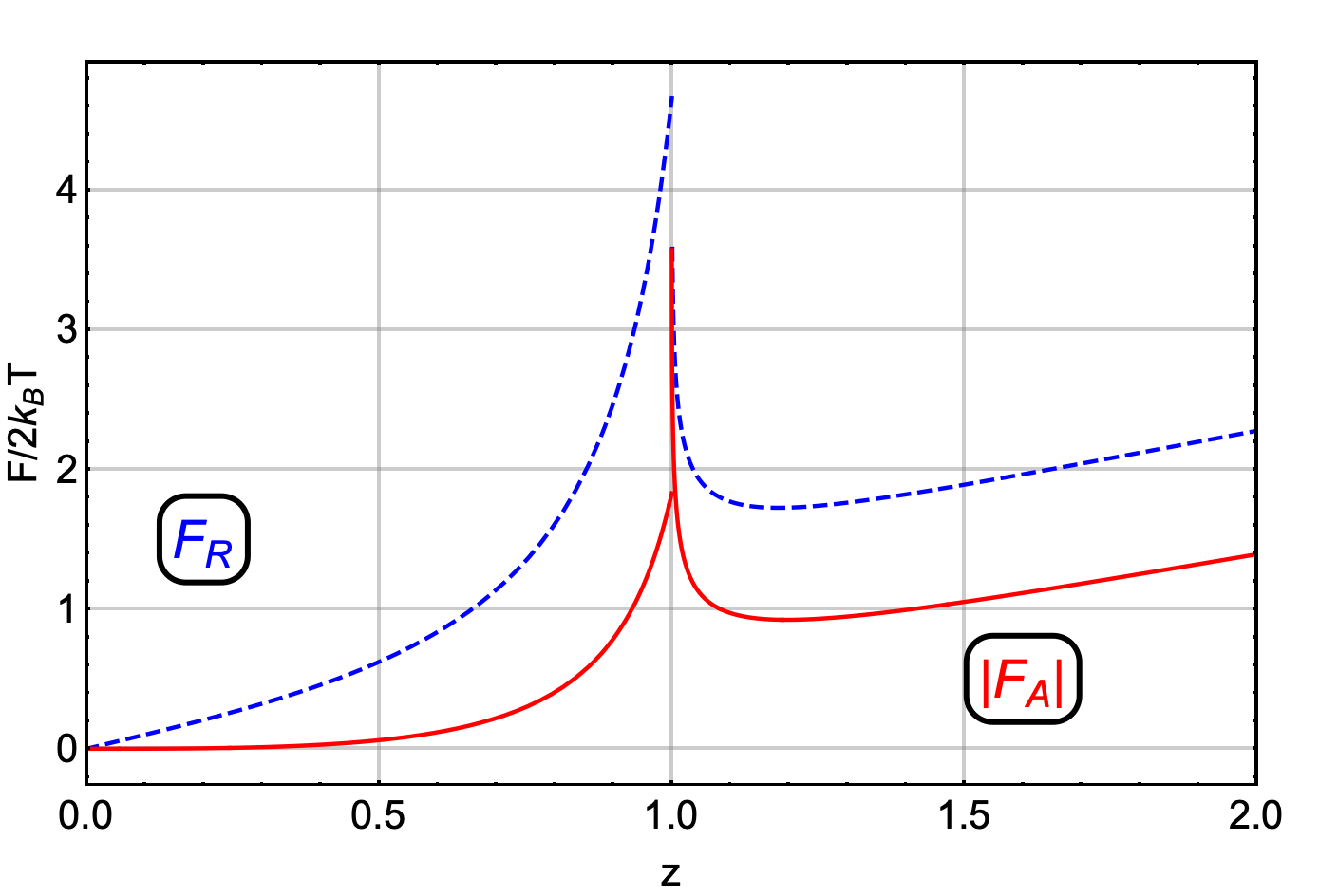}
\caption{Magnitude of the repulsive, $F_R$, and attractive, $F_A$, forces in the case of two disks inside a box with periodic boundary conditions corresponding to Eqs. (\ref{eq:2diskForceB}) and  (\ref{eq:2diskForceA}), respectively. Notice that $\vert F_R \vert > \vert F_A \vert$ for all values of $z$.} \label{fig:forces2Disk}
\end{figure}

In Fig. \ref{fig:forces2Disk} we have plotted the attractive and repulsive forces.  Clearly, the repulsive force always wins over the attractive one.

Now, in the case of  two disks in a  square box with {\it non-periodic boundary conditions}, then $A\rightarrow A_c=A(1-f_0)$. In this case the force is
\begin{equation}
F_A=
\begin{cases}
-\frac{2k_B T \sigma^3}{\Omega (2,A_c(z))} \left(\pi + \frac{24}{z^2} \right) \; , \text{for} \; z<1\; . \\ \\
 -\frac{2k_B T \sigma^3}{\Omega (2,A_c(z))} \left(4 \sec^{-1}(z)+\frac{24}{z^2}+\frac{4}{\sqrt{z^2-1}} \right. \\
 \left. + \frac{16}{z^3 \sqrt{z^2-1}} \right)\; , \text{for} \; z>1\; .
\end{cases}
\end{equation}
and 
\begin{equation}
F_R=
\begin{cases}
\frac{2k_B T \sigma^3}{\Omega (2,A_c)} \left(\frac{(\pi-1) z}{2}+\frac{10}{z}+\frac{16}{z^3} \right)  ,\; z<1\; . \\ \\
 \frac{2k_B T \sigma^3}{\Omega(2,A_c(z))} \left(\pi-2+\frac{16}{z^3}+\frac{12-2\pi}{z}+\frac{8 \sec^{-1}(z)}{z} \right. \\
\left.\qquad +\frac{12}{z^2\sqrt{z^2-1}}+\frac{6}{z \sqrt{z^2-1}} \right), z>1\; .
\end{cases}
\end{equation}

\begin{figure}[htp]
\centering
\includegraphics[width=2.3in]{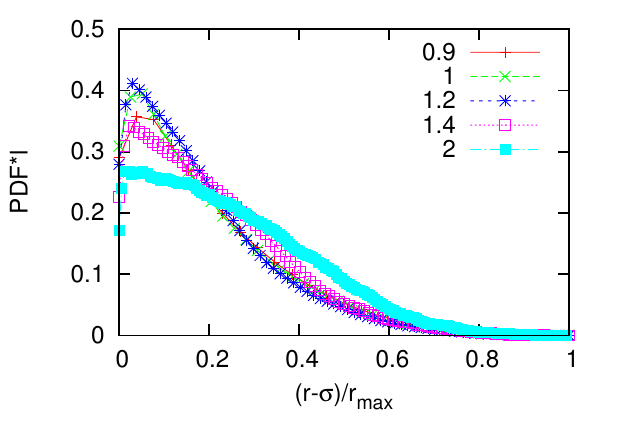}
\includegraphics[width=2.3in]{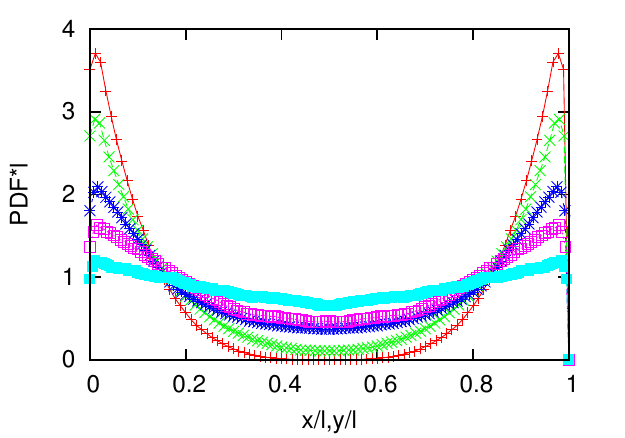}
\caption{\label{two} Distribution of lengths between particles (left) and between particles with respect to the walls (right) for two disks inside a non-periodic square box in reduced units. The different curves correspond to $l=L-\sigma=0.9,1,1.2,1.4,2$. The diameters of the disks were kept fixed at $\sigma=1$. As the length of the box decreases, there are some configurations more probable than others and the PDF tends to a $\delta$-function. On the contrary, as the length of the box increases, the PDF becomes rather flat and all configurations are equiprobable. 
}
\label{Fig:2DisksPDF}
\end{figure}

In Fig. \ref{Fig:2DisksPDF} we present the PDF of the distance between disks as well as the distance between disk and walls for the case of a two disks inside a non-periodic square box. Notice how for a large box the distributions are rather flat and each distance configuration has essentially the same probability. However, as the box decreases in size, i.e., as the overlaps increase there are some distances configurations more probable than others. Eventually, as $L$ decreases, the distance between disks' PDF becomes a Dirac delta function $\delta(r-\sigma)$, while the distance between disks and walls becomes two Dirac delta functions because each disk gets stuck in a corner of the box.

\section{Second-neighbors excluded volume overlap in a five particles in a box system}

\begin{figure*}[hbtp] 
	\centering
	a)\includegraphics[scale=0.20]{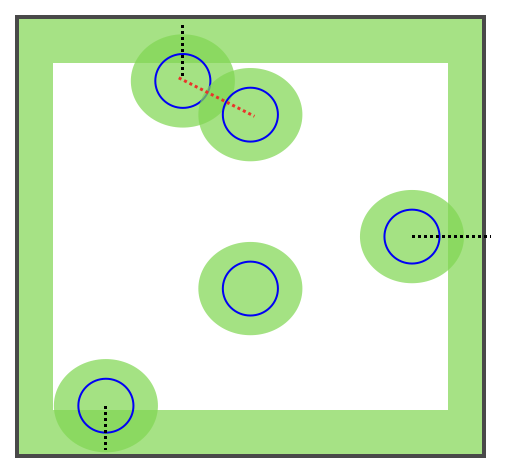}
	b)\includegraphics[scale=0.20]{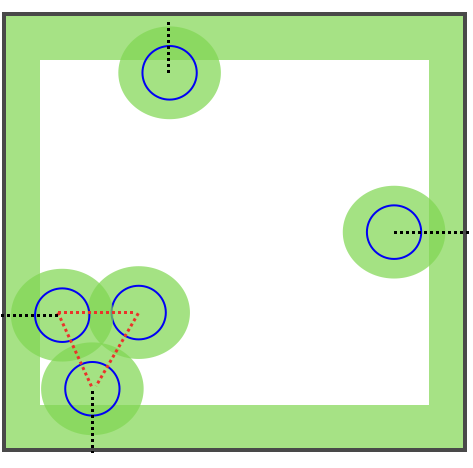}
	c)\includegraphics[scale=0.16]{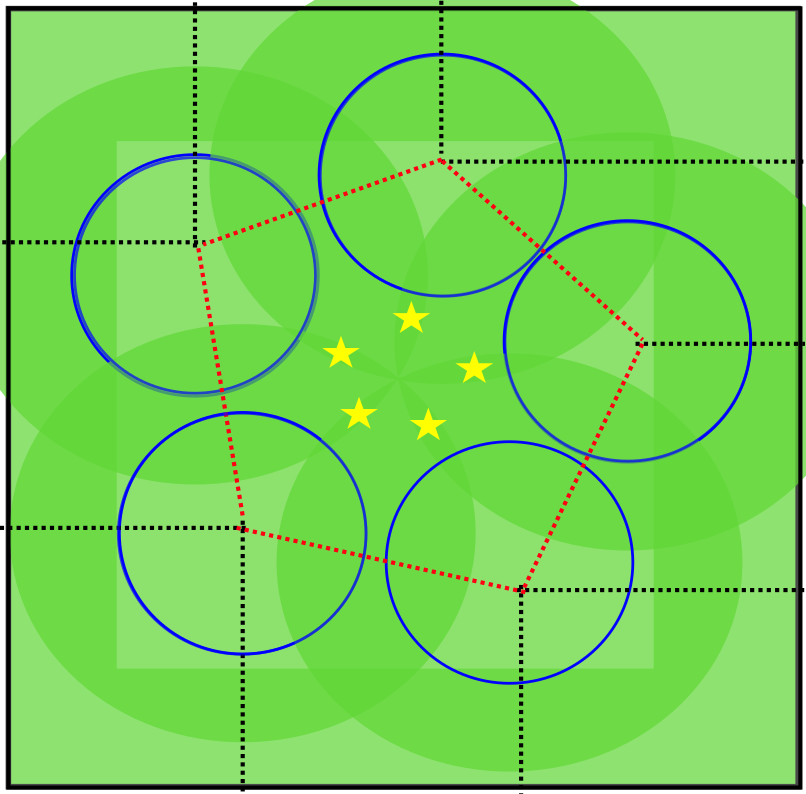}
	d)\includegraphics[scale=0.21]{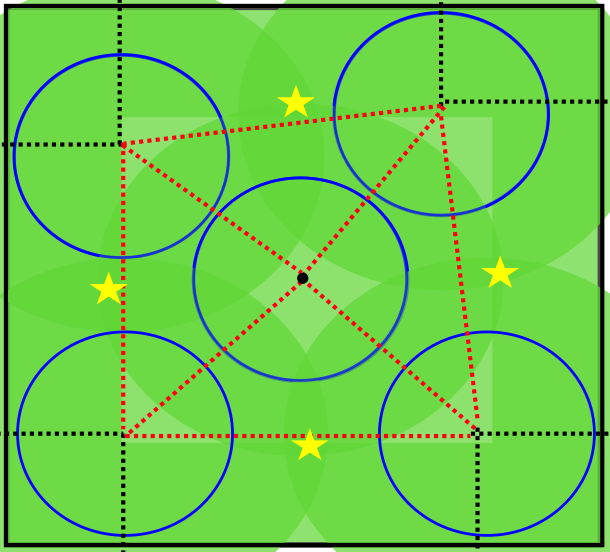}
	\caption{Sketch of five hard-disks in a box. In blue we show the diameter $\sigma$ of the hard-disks, in green we show the excluded volume of each hard-disk as well as the excluded volume with the hard walls of the box. The overlap of excluded volumes correspond to a darker green and are marked with dashed lines of different colors to further make it clear when this occurs. The red dashed line corresponds to the overlap of excluded volumes between hard-disks, whereas the black dashed lines correspond to the overlap of excluded volumes between a wall and a hard-disk. The stars mark the excluded volume overlap between second neighbors.  Panels a)-b) the sizes of the box corresponds to to $L=10\sigma$. Panels c) and d) correspond to  $L=4.8 \sigma$ }\label{Fig:snapshot} 
\end{figure*}

Let us now consider the
simple case of five particles in a box, a system that has been investigated numerically \cite{bowles1999five}. As explained at the introduction of this work, this system has the possibility to crystallize as well as to rearrange in an amorphous phase \cite{bowles1999five}.  This system has a very simple energy landscape diagram. In particular, we will numerically show that the basins of the energy landscape are determined by the second-neighbor excluded volume overlap. Here we use the Monte Carlo simulations method to sample the configurations of the five hard disk in a box model. Figure \ref{Fig:snapshot} shows a collection of snapshots depicting the simulation of five hard disks particles in a square box.

 In the same way as in the previous systems, we have drawn the hard disks in blue, the excluded volume of each hard disk in green. In all cases, the excluded volumes due to the walls are also colored in green. In Figure \ref{Fig:snapshot} we indicate by dashed red lines when an excluded-volume overlap appears between two hard disks. The dashed black lines represent an excluded volume overlap of a wall with a disk. Thus, these lines represent the emergence of attractive and repulsive forces as the ones shown in figures \ref{fig:forces1Disk} and \ref{fig:forces2Disk}.
  We can notice that at low densities the overlap of excluded volume, both with walls and hard disks is not very probable. However, as the density increases, the probability of excluded volume  overlap increases. Such overlaps act as restrictions against the rest of the hard disks. 

 From  Figure \ref{Fig:snapshot}, we can notice that the free volume decreases to a point in which a collective motion (in this case a rotation) is necessary to sample all available free volume. Actually, we can see from the evolution in Figure \ref{Fig:snapshot} panel b) to c), that overlap percolation with walls, and particles have occurred, as in the Hoover large system the free volume becomes an intensive quantity.  Another other important remark is that the overlap of excluded volume occurs with the second nearest neighbors and walls. For example, in panel d) of Fig. \ref{Fig:snapshot} the central particle excluded volume  overlaps with all the walls, restricting the motion of all other particles to pass through the central particle and the walls. In c) we can observe the overlap of all the particles with the second neighbors restricting the mobility of each particle towards the center of the system. This second-neighbor excluded volume percolation precisely coincides with the inherent structure basin of the energy landscape, which was determined by Bowles and Speedy \cite{bowles1999five}. In fact, it has been shown using
 non-linear optimization theory, that the inherent structures of the energy landscape are given by the number of constraints \cite{NaumisPRE2005} which in this case are given by the  red lines in Fig. \ref{Fig:snapshot}. 

\begin{figure*}[htp]
\centering
a)\includegraphics[scale=1]{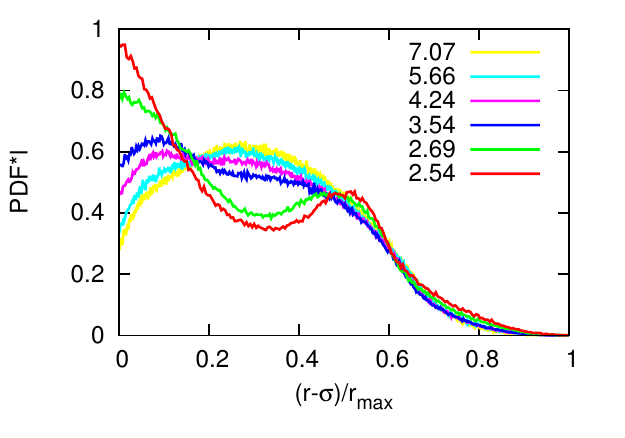}\includegraphics[scale=1]{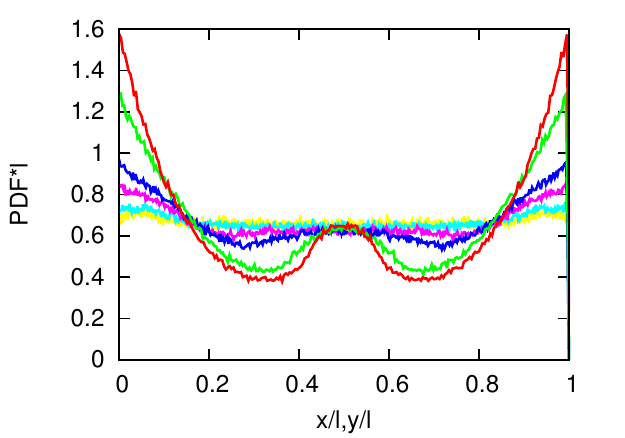}
b)\includegraphics[scale=1]{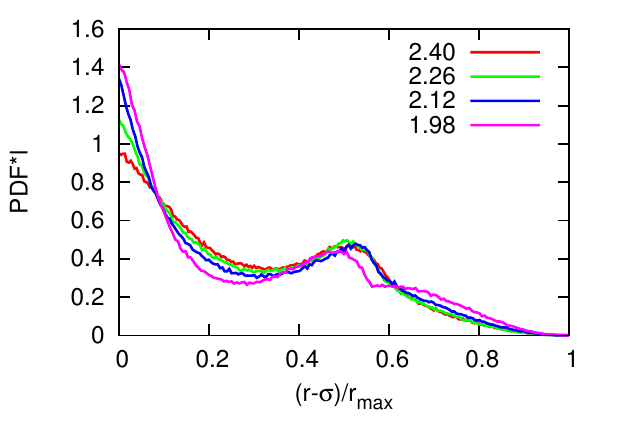}\includegraphics[scale=1]{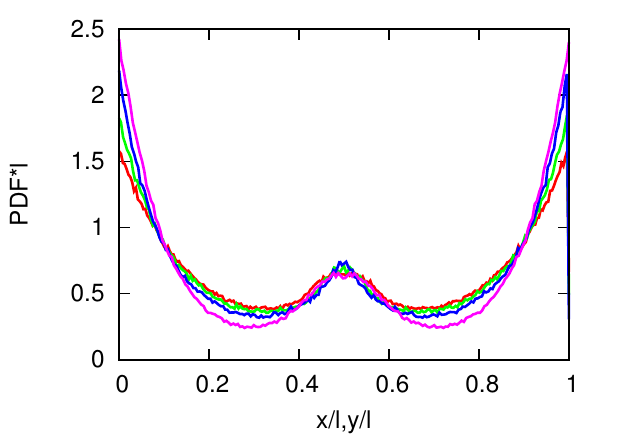}
c)\includegraphics[scale=1]{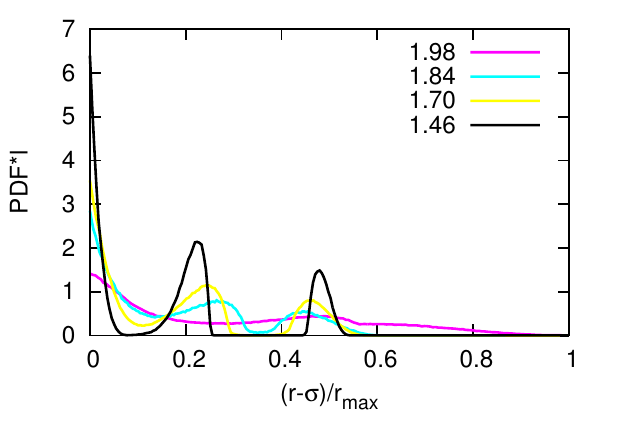}\includegraphics[scale=1]{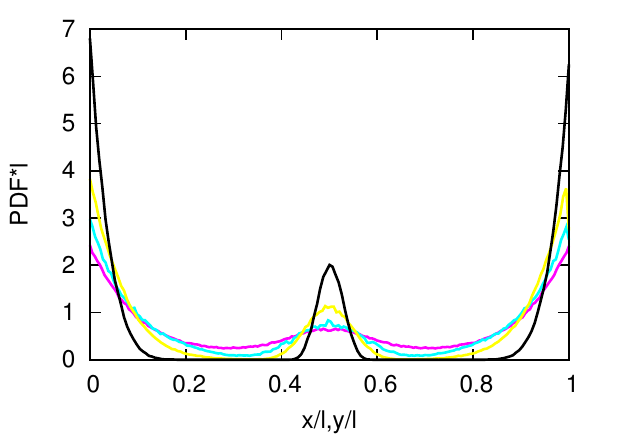}

\caption{\label{fig2} Distribution of lengths  between particles (left column) and between particles with respect to the walls (right column) in reduced units of the maximally observed distance ($r_{max}$). In row a), the different curves are for box lengths $l=L-\sigma=7.07,5.66,4.24,3.54,2.69,2.54$. These values correspond to the diluted limit. In panel b), the curves are for $l=L-\sigma=2.40,2.26,2.12,1.98$. Finally, in panel c) we present curves for $l=L-\sigma=1.98,1.84,1.70,1.46$. These values correspond to the dense limit. In all cases, $\sigma=1$. Notice how a shoulder appears in left panel a), and hoe the basins of the landscape are built, with a central peak indicating a central disk, and a confinement for the other four disks with the walls and a central disk.}

\end{figure*}

In order to understand the role of the restrictions formation and how it contributes to the stabilization of a basin in the configurational energy landscape, we next calculate the distance distributions between the center of the hard disks and also each particle with the walls of the hard box. In figure \ref{fig2} we show both distributions and the total distribution. Notice in Fig. \ref{fig2} the formation of a shoulder for distances less than two diameters, representing the overlap of the excluded volume, similar to the ones observed in previous works \cite{huerta2006freezing,huerta2015collective}.

\section{Conclusions}

In this work we have analyzed the role of  excluded volume overlap in the thermodynamics of simple systems. This allows to find analytical expressions for the equation of state and attractive/repulsive forces for one and two disks in a box. The case of one disk is interesting as is an useful approximation to a very dense fluid or a solid. The attractive force increases with the packing and is independent upon the kind of polygonal box. In fact, this simple analysis allows to identify in a rough way the range in which freezing will occur.
For two disks, the attractive and repulsive forces have a peak.  For all packings, the attractive force remains less than the repulsive one. 
For the five disk in a box system, we numerically showed that the basins of the energy landscape are determined by the second-neighbor excluded volume overlap, a result that can only be explained in terms of the attractive and repulsive effective forces resulting from entropic forces, instead of direct collisions.
Therefore, this result suggests that the role of the constraints due to the overlap of the excluded volume of second-neighbor interactions in defining basins in the configuration energy landscape.

\section{Acknowlegments}
AH thanks the support of CONACyT during the sabbatical leave and valuable discussions with Dra. Karen Volke and Dr. Alejandro Vasquez of the IFUNAM; also AH thanks 
to Dr. Andrij Trokhymchuk and Taras Bryk of the ICMP of the National Academy of Sciences Ukraine for valuable discussions. This work has been supported by DGAPA-UNAM project IN102717. J.Q.
Toledo acknowledges CONACyT for a PhD scholarship.

\section{References}

\bibliography{mybib}

\end{document}